\documentclass{vgtc}                          

\ifpdf
  \pdfoutput=1\relax                   
  \usepackage{graphicx}                
  \DeclareGraphicsExtensions{.pdf,.png,.jpg,.jpeg} 
\else
  \usepackage{graphicx}                
  \DeclareGraphicsExtensions{.eps}     
\fi%

\graphicspath{{figures/}{pictures/}{images/}{./}} 

\usepackage{microtype}                 
\PassOptionsToPackage{warn}{textcomp}  
\usepackage{textcomp}                  
\usepackage{mathptmx}                  
\usepackage{times}                     
\usepackage{cite}                      
\usepackage{tabu}                      
\usepackage{booktabs}                  

\onlineid{0}

\vgtccategory{Research}

\vgtcinsertpkg

\title{Never Tell the Trick:\\
Covert Interactive Mixed Reality System for Immersive Storytelling}

\author{Chanwoo Lee\thanks{e-mail: chanwoo.lee@imperial.ac.uk}\\ %
        \parbox{1.6in}{\scriptsize \centering Imperial College London \\ Royal College of Art\\ Multimodal XR Lab, Modulabs} %
\and Kyubeom Shim\thanks{e-mail: rbqja0415@naver.com}\\ %
     \parbox{1.6in}{\scriptsize\centering  Dept. of Art \& Technology, \\ Sogang University}%
\and Sanggyo Seo\thanks{e-mail: ssk7808@gmail.com}\\ %
     \parbox{1.6in}{\scriptsize \centering Dept. of Art \& Technology, Sogang University}
 \and Gwonu Ryu\thanks{e-mail: gwonu@hyundai.com}\\ %
     \parbox{1.6in}{\scriptsize \centering Hyundai Motor Group}
 \and Yongsoon Choi\thanks{e-mail: yongsoon@sogang.ac.kr}\\ %
     \parbox{1.6in}{\scriptsize \centering Dept. of Art \& Technology, Sogang University}}

\teaser{
\centering
 \includegraphics[alt={Overview of three scenes with performers interacting in a game setup. Left: A performer with a lit cube prop in a room with a projected natural landscape. Center: A lone performer in a space with industrial-style lighting and metallic structures. Right: A performer adjusting a cube on a reflective floor in a room with a virtual forest scene.}, width=0.9\textwidth]{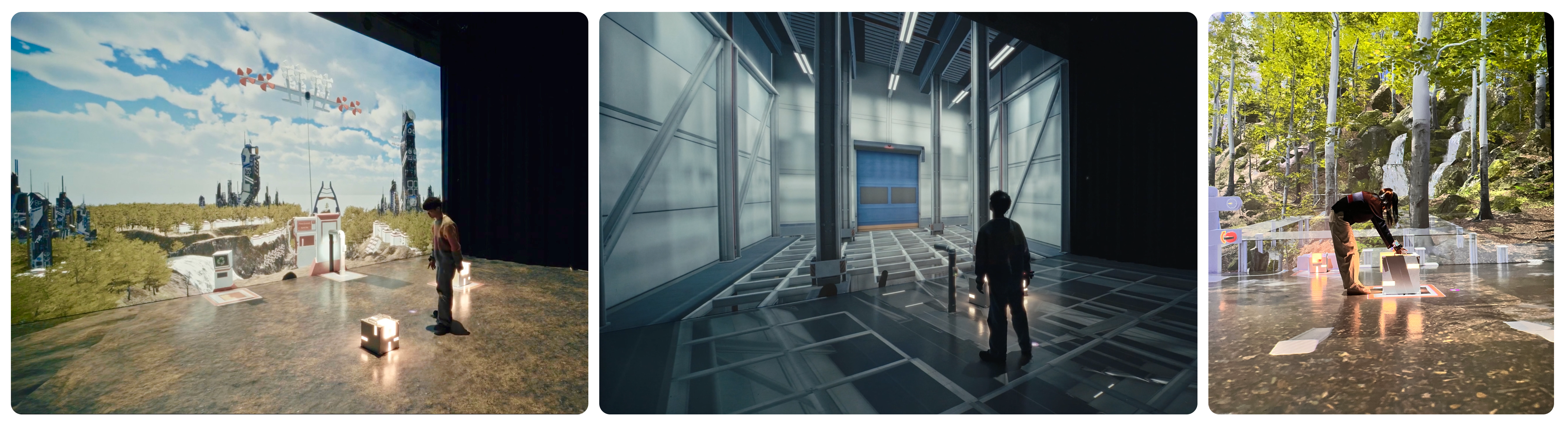}
 \caption{Overview of immersive gameplay in MR environment}
}

\abstract{
This study explores the integration of Ultra-Wideband (UWB) technology into Mixed Reality (MR) Systems for immersive storytelling. Addressing the limitations of existing technologies like Microsoft Kinect and HTC Vive, the research focuses on overcoming challenges in robustness to occlusion, tracking volume, and cost efficiency in props tracking. Utilizing UWB technology, the interactive MR system enhances the scope of performance art by enabling larger tracking areas, more reliable and cheaper multi-prop tracking, and reducing occlusion issues. Preliminary user tests suggest meaningful improvements in immersive experience, promising a new possibility in Extended Reality (XR) theater, performance art and immersive game.
} 

\CCScatlist{
  \CCScatTwelve{Human-centered computing}{Human computer interaction (HCI)}{Interaction paradigms}{Mixed / augmented reality};
  \CCScatTwelve{Human-centered computing}{Ubiquitous and mobile computing}{Ubiquitous and mobile computing tools}{};
  \CCScatTwelve{Applied computing}{Arts and humanities}{Media arts}{}
}

\begin{document}


\firstsection{Introduction}

\maketitle

In the context of immersive content, such as theater and performance art, losing awareness of the `real' world is a critical aspect of immersion. \cite{lidwell2010universal}
This is achieved by seamlessly integrating technology within the performance.

Immersive theater are characterized by expansive stage, installations for scenic design, performers in creative costumes navigating complex paths, and the use of multiple props. Although technologies based on the Time-of-Flight (ToF) principle using infrared light like Microsoft Kinect and HTC Vive  have been employed in this field \cite{kurillo_evaluating_2022}, they face three primary challenges for immersive theater:

\textbf{Robustness to Occlusion:} Sensors can not be concealed in props or costumes. The abundance of stage installations for scenic design limits the movement of actors, as they cannot be tracked when walking or hiding behind them\cite{kurillo_evaluating_2022}. Furthermore, Kinect struggles when actors are wearing unconventional costumes\cite{xia_human_2011}.

\textbf{Tracking Volume:} The tracking range of current solutions is limited, restricting performers' space \cite{bowman_questioning_2012} - Azure Kinect has a maximum of 3.5m\cite{kurillo_evaluating_2022}, and Vive Lighthouse 2 covers 100m²\cite{noauthor_steamvr_nodate}.

\textbf{Cost of Tracking Multiple Props:} The high cost of tracking devices makes tracking numerous props financially burdensome.

Meanwhile, Ultra-Wideband (UWB) technology emerges as a promising alternative for large-scale tracking \cite{dardari_indoor_2015}. This poster aims to broaden the expressive capabilities of immersive theater by embedding UWB technology into Interactive Mixed Reality (MR) Systems. 

\section{Implementation}

\subsection{Immersive Stage Design}


The performance stage, featuring MR systems, spans an area of 1042cm in width and 1044cm in depth, affording ample space for performers, which is vital for natural interaction \cite{bowman_questioning_2012}. It is equipped with two 4K projectors for front and floor projection - Christie Mirage 304K and Boxer 4K30, both offering brightness of 30,000 center lumens, ensuring a vibrant and immersive experience. 

\subsection{Tracking System Design}

\begin{figure}[h]
    \centering
    \includegraphics[alt={An image of two illuminated cubic performance props and one taller, cylindrical prop with a red light on middle on a darkened stage.}, width=0.6\linewidth]{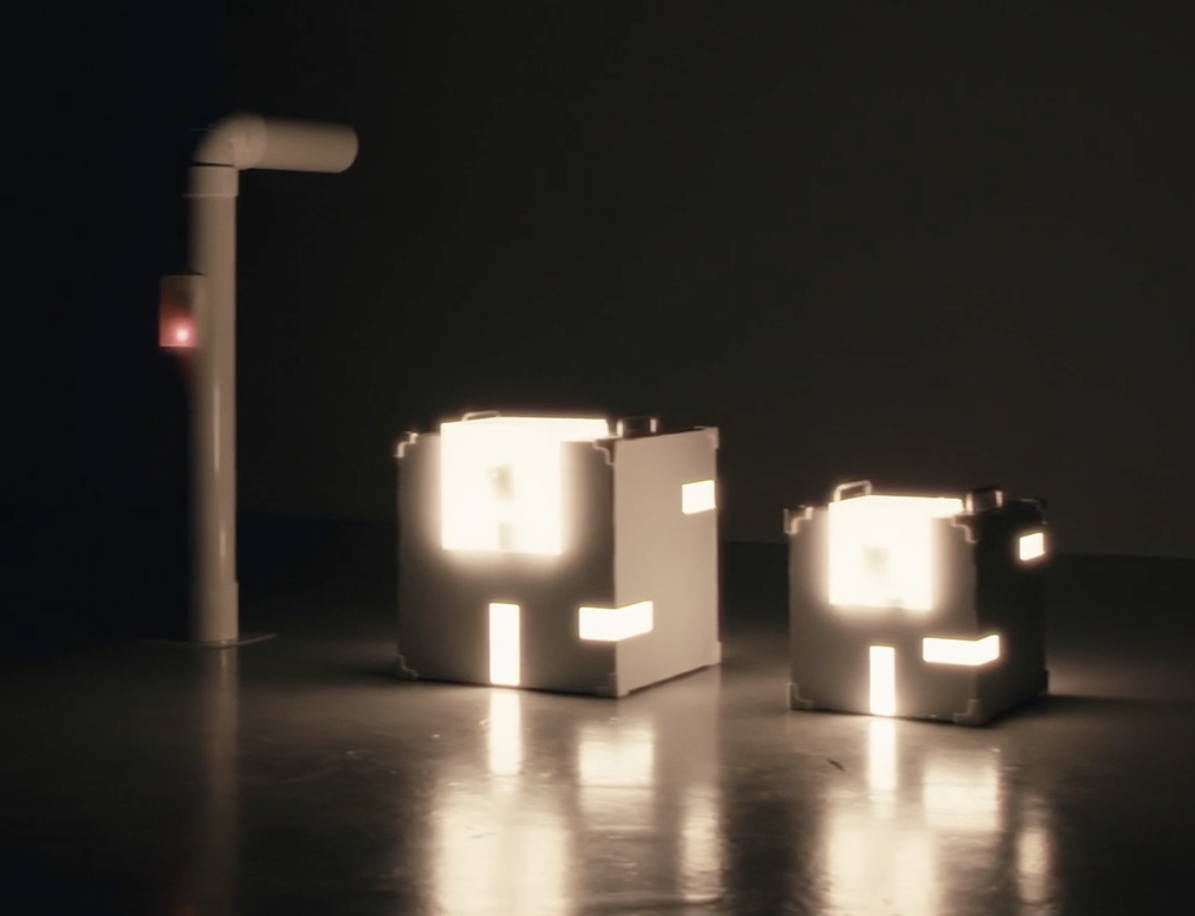}
    \caption{3 performance props with concealed UWB sensors}
    \label{fig:enter-label}
\end{figure}

The UWB system was constructed using the Pozyx Creator Kit, priced at 1350 Euros, including 4 fiexed (Anchor) and 6 rover (Tag) transreceivers \cite{noauthor_creator_nodate}. This setup presents a cost advantage compared to assembling an equivalent system with Vive Tracker 3.0 units, priced at 154 Euros each, and Base Station 2.0 units, at 245 Euros each, which together would total 1904 Euros \cite{noauthor_steamvr_nodate}.

4 anchors are installed on the batten of the stage, and one main tag is connected to the main computer via arduino uno and three tags are in the props. The main tag gathers location of three tags and sends the data to the Unity via serial communication \cite{dabove_indoor_2018}. This 4 anchor system can cover up to 400 – 800m² space \cite{noauthor_creator_nodate}.

\begin{figure}[h]
    \centering
    \includegraphics[alt={Figure depicts three stages of Ultra-Wideband (UWB) anchor setup: a) Measurement with a laser, b) Installation of UWB anchors on a batten, c) Testing data output on a laptop connected to a uwb tag.}, width=1\linewidth]{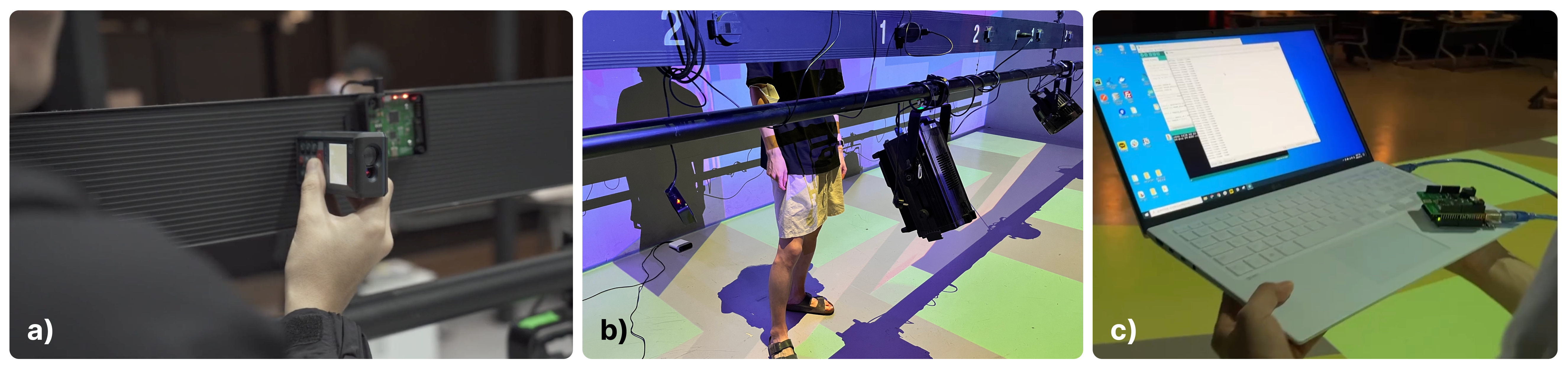}
    \caption{Process of measuring, installing and testing the anchors}
    \label{fig:enter-label}
\end{figure}

While installing and configuring the system, four anchors at a distance of three meters in a square configuration didn't cover all 10m*10m area. Throughout the trial and failure, the tracking area was improved to fit the environment by reconfiguring the anchors to form a 755cm*570cm rectanlge. 

Through a bidirectional ranging strategy, it measures signal ToF between the TAG and ANCHORs, accurately calculating the TAG's position. An inbuilt Inertial Measurement Unit (IMU), which includes accelerometers, gyros, and magnetometers, which contribute to a positional accuracy within 320±30mm and easily can become lower to 100±25mm \cite{dabove_indoor_2018}.

\subsection{Application}

\begin{figure} [h]
    \centering
    \includegraphics[alt={An image of a) A cube augmented with projections with a person standing nearby. b) A diagram showing the recognition boundary around the cube with inner dimension labeled as 40cm and outer dimention by 65cm.}, width=0.9\linewidth]{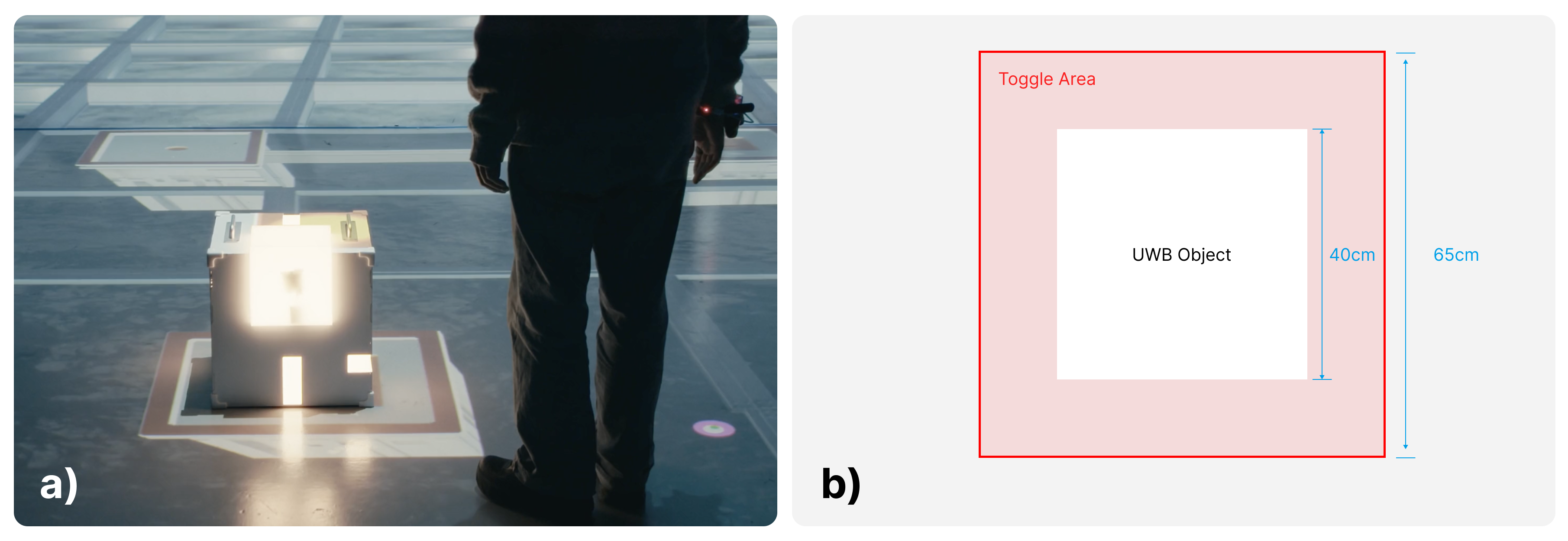}
    \caption{Cube augmented with projection and recognition boundary}
    \label{fig:enter-label}
\end{figure}
For the system application, interactive MR performance involving puzzle element was created with 3 props - or `cubes' and using Unity. Each cube contains a Pozyx tag inside the acrylics and wooden body for position tracking. Contrast to the vive tracker, these tags were able to be hidden inside the props thanks to the UWB  technology.

Being physically present in the stage, props foster a sense of spatial immersion by allowing users to navigate the expansive environment. These cubes also serve as key elements of a puzzle, with their placement within designated graphics areas triggering progression to the next stage or scene. 

To preserve location data reliability, location toggling occurs only if the TAG remains within the area for a minimum of 100 frames. Given the cube dimensions (the largest being 40*40*40cm) and Pozyx's positional accuracy (ranging from about 100-300mm \cite{dabove_indoor_2018}), a toggle location area of 65*65 cm is designated for each cube.

The system and content were experienced by 54 individuals who successfully completed a 15-minute play session. Among them, 45 are students and alumni of the Art \& Technology program at Sogang University, aged between 20 and 29. Their feedback highlighted the enhanced immersion (`I didn’t know how it worked, I was just amazed'), interactivity (`It was so natural and amazing to see the graphics change according to my perspective and to move the box to progress to the next stage'), and naturalism (`I usually feel motion sick in VR, but it was nice to not have motion sickness while actually walking'). These improvements in 3D environments are achieved thanks to fewer limitations than traditional tracking systems. This information was translated from Korean using Google Translate.

\section{Conclusion and Future Work}
An Immersive MR System for a 10m*10m stage and performance was devised, incorporating UWB sensors for the Robustness to Occlusion, Increased Tracking Volume and Low Cost of Multiple Props Tracking. 

The system and content have been tested and played by 54 users who participated as performers, successfully completing the game experience from start to finish. This validated the concept of using UWB for tracking within MR systems.

The findings highlight the advantages of UWB: expanded tracking range, robustness to occlusion and multi-object tracking, opens the door for dynamic, collaborative, and expressive theater and performance. This is a crucial part in fields like immersive storytelling, XR theater and performance, enabling unrestrained creative expression in areas such as stage size, scenic design, costume, actor path, etc. The technology supports multi-user tracking without occlusion and utilizes a large space, effectively eliminating spatial constraints.

Through such a covert interactive MR system, the range of expressions possible in immersive theater has expanded, leading to the potential emergence of new and more immersive and seamless content, especially in the field of immersive game and XR theater.

\acknowledgments{
This work was supported by the National Research Foundation of Korea, funded by the Ministry of Education, under the `University Innovation' program. This research was supported by Brian Impact Foundation, a non-profit organization dedicated to the advancement of science and technology for all.

We wish to express our thanks to Taehyun Kim for his contribution to the 3D models in the virtual world.}

\bibliographystyle{abbrv-doi}

\bibliography{template}
\end{document}